\documentclass{article}
\usepackage{arxiv}

\usepackage{amssymb}

\usepackage{listings, algorithm, amsmath, algpseudocode, tikz, xspace, nameref, caption, url}
\usepackage[color=yellow]{todonotes}
\usetikzlibrary{calc, matrix, positioning, automata, positioning, shapes, shapes.geometric, arrows, decorations.pathmorphing, plotmarks, trees, fit}
\newcommand{\highlightcolour}{white}
\usepackage{color, xcolor, soul}
\colorlet{customhighlight}{\highlightcolour}
\sethlcolor{customhighlight}

\makeatletter
\renewcommand{\Function}[2]{%
  \csname ALG@cmd@\ALG@L @Function\endcsname{#1}{#2}%
  \def\jayden@currentfunction{#1}%
}
\newcommand{\funclabel}[1]{%
  \@bsphack
  \protected@write\@auxout{}{%
    \string\newlabel{#1}{{\jayden@currentfunction}{\thepage}}%
  }%
  \@esphack
}
\makeatother

\newcommand{\consult}{\textsc{Consult}\xspace}
\newcommand{\mathcolourbox}[2]{\colorbox{\highlightcolour}{$\displaystyle #2$}}
\makeatletter
\newcommand{\algcolor}[2]{%
  \hskip-\ALG@thistlm\colorbox{#1}{\parbox{\dimexpr\linewidth-2\fboxsep}{\hskip\ALG@thistlm\relax #2}}%
}
\makeatother



\begin{document}

\title{A semi-autonomous approach to connecting proprietary EHR standards to FHIR}

\author{
  Martin Chapman\thanks{*corresponding author} \\
  Department of Population Health Sciences \\ King's College London \\ 
  United Kingdom \\
  \texttt{martin.chapman@kcl.ac.uk} \\
  \And
  Vasa Curcin \\
  Department of Population Health Sciences \\ 
  King's College London \\ 
  United Kingdom \\
  \texttt{vasa.curcin@kcl.ac.uk} \\
  \And
  Elizabeth I Sklar \\
  Department of Informatics \\ 
  King's College London \\ 
  United Kingdom \\
  \texttt{elizabeth.sklar@kcl.ac.uk} \\
}

\maketitle 

\begin{abstract}
HL7's Fast Healthcare Interoperability Resources (FHIR) standard is designed to provide a consistent way in which to represent and exchange healthcare data, such as electronic health records (EHRs). SMART--on--FHIR (SoF) technology uses this standard to augment existing healthcare data systems with a standard FHIR interface.
While this is an important goal, little attention has been paid to developing mechanisms that convert EHR data structured using proprietary schema to the FHIR standard, in order to be served by such an interface. 
In this paper, a formal process is proposed that both identifies a set of FHIR resources that best capture the elements of an EHR, and transitions the contents of that EHR to FHIR, with a view to supporting the operation of SoF containers, and the wider interoperability of health records with the FHIR standard.
This process relies on a number of techniques that enable us to understand when two terms are equivalent, in particular a set of similarity metrics, which are combined along with a series of parameters in order to enable the approach to be tuned to the different EHR standards encountered.
Thus, when realised in software, the translation process is semi-autonomous, requiring only the specification of these parameters before performing an arbitrary number of future conversions. 
The approach is demonstrated by utilising it as part of the \consult project, a wider decision support system that aims to provide intelligent decision support for stroke patients.

\end{abstract}

D2.1 (Software Engineering) Requirements/specification J.3 (Life and Medical Sciences) Fast Healthcare Interoperability Resources (FHIR) SMART--on--FHIR Health data interoperability Decision support systems




\section{Introduction}\label{sec:introduction}

The increased interest in health data, ranging from \textit{electronic health records} (EHRs), via wearable devices to large research data repositories, has resulted in an explosion of software tools that interact with these data sources and provide personalised portals, decision support based on machine learning models, trial recruitment assistance and other functions. Standardisation of access to these data sources \hl{can be implemented at two levels}, that of a common data model, typically used for consistent representation of research data repositories and exemplified by standards such as OMOP CDM \cite{MarcOverhage2012}, CDISC SDTM \cite{Kuchinke2009}, PCORNET \cite{Collins2014} and the Sentinel model, and that of a messaging protocol such as HL7 FHIR, which is used to request individual data elements from the data source \cite{Bender2013}. The differences between the two stem from their different purposes: the former aim to preserve the richness of the collected data, and to optimise the storage for large data collections through normalisation; the latter focus on producing self-contained messages for the client requesting some information. While some overlap is possible, in essence, common data models are typically determined by the producer, while messaging protocols are directed at consumer needs.

This producer-consumer dichotomy comes to the fore in situations where a consumer may require a FHIR \textit{Application Programming Interface} (API), but is only presented with a proprietary interface to the internal data model, which could be excessively rich and complex. For example, when an external software tool (consumer) require access to the API of an EHR system (producer), in order to perform queries and make updates for the purposes of decision support or clinical trial recruitment, it is expected to reason with payloads of data formatted according to the EHR vendor's internal data model. 

In addition to the complexity exhibited by the internal data models offered by EHR vendors, it is often that case that vendors adopt different internal data models. Even if they aim to adopt the same standard, discrepancies are still likely to be present, e.g. through the version of the standard used. This presents a difficulty for those designing pieces of client software that require access to data from a variety of different vendors, as the data received must be parsed differently for each standard. This situation is present in the \textit{\consult} decision support system~\cite{Kokciyan2018}, which aims to autonomously generate treatment plans for stroke patients by gathering data about those patients from a variety of sources, including EHRs (Figure \ref{fig:consult-pre}). The \consult prototype uses patient data from SystmOne (TPP) and mMedica (Asseco) primary care EHR systems, which can be installed on machines in two remote locations and source patient data in different formats. Currently, if the \consult system needs to reason within data from both of these EHRs, it has to do so for each (complex) vendor format separately.

\begin{figure}[]
\begin{center}
\centering\scalebox{0.7}{
\tikzset{rectangle state/.style={draw,rectangle}}
\begin{tikzpicture}[node distance=3.5cm,initial text=,
    database/.style={
      cylinder,
      shape border rotate=90,
      aspect=0.25,
      draw
    }]
\begin{scope}
\node[rectangle,very thick,minimum size=10mm, align=center] (consult) {\consult \\ system};
\node[rectangle state,very thick,minimum size=10mm, align=center] (so-module) [above left of=consult]  {SystmOne \\ module};
\node[rectangle state,very thick,minimum size=10mm, align=center] (so-installation) [left of=so-module]  {SystmOne \\ installation};
\node[database,very thick,minimum size=10mm, align=center] (so-data) [left of=so-installation] {SystmOne \\ formatted \\ data};
\node[rectangle state,very thick,minimum size=10mm, align=center] (m-module) [below left of=consult] {mMedica \\ module};
\node[rectangle state,very thick,minimum size=10mm, align=center] (m-installation) [left of=m-module] {mMedica \\ installation};
\node[database,very thick,minimum size=10mm, align=center] (m-data) [left of=m-installation] {mMedica \\ formatted \\ data};
\node[rectangle state,very thick,minimum size=10mm, align=center] (process-so) [above right of=consult]{Process \\ SystmOne data};
\node[rectangle state,very thick,minimum size=10mm, align=center] (process-m) [below right of=consult]{Process \\ mMedica data};
\node[rectangle state,very thick,minimum size=10mm, align=center] (reasoner-a) [right of=process-so]{Reasoner A};
\node[rectangle state,very thick,minimum size=10mm, align=center] (reasoner-b) [right of=process-m]{Reasoner B};

\end{scope}
\begin{scope}
\path[<->] (consult) edge [bend right,very thick] node [left] {} (so-module);
\path[<->] (so-module) edge [very thick] node [above] {Tunnel} (so-installation);
\path[<->] (consult) edge [bend left,very thick] node [left] {} (m-module);
\path[<->] (m-module) edge [very thick] node [below] {Tunnel} (m-installation);
\path[->] (so-installation) edge [very thick] node [] {} (so-data);
\path[->] (m-installation) edge [very thick] node [] {} (m-data);
\path[->] (consult) edge [very thick] node [] {} (process-so);
\path[->] (consult) edge [very thick] node [] {} (process-m);
\path[->] (process-so) edge [very thick] node [] {} (reasoner-a);
\path[->] (process-m) edge [very thick] node [] {} (reasoner-b);
\end{scope}
\end{tikzpicture}}

\caption{The \consult prototype.}
\label{fig:consult-pre}
\end{center}
\end{figure}

To address these issues, it has been suggested that a vendor's internal data model should be \textit{connected} to a FHIR interface using technology such as \textit{SMART--on--FHIR containers}, which encapsulate a system such as an EHR and expose a FHIR-compliant interface to applications that wish to interact with the original system~~\cite{Mandel2016}. However, the challenge of serving data in the FHIR standard, through a SMART--on--FHIR container, despite an arbitrary standard being used by the encapsulated system, remains. A typical approach is to identify which elements of the FHIR standard are considered \textit{equivalent} to elements of the proprietary EHR schema when used to annotate data~\cite{Wagholikar2017}. Then, any data marked up using the schema can be \textit{mapped} into the FHIR standard (e.g. by replacing element tag names with their FHIR equivalent), prior to be served by the container. Typically, equivalent FHIR elements are specified manually. However, this approach assumes the \hl{EHR} schema is available, which is unlikely to be the case, especially when working with global EHR integrations or legacy EHR software. Even if the schema is available, EHR software schema can contain a significant number of distinct elements -- in excess of 100 for SystmOne -- which are often arranged into complex hierarchies. Once identified, each of these elements must then be compared with the elements of FHIR, which itself contains in excess of 100 resources. As such, specifying equivalent FHIR elements manually requires knowledge of both the EHR schema and of FHIR, is time-consuming and does not scale. In addition, statically identified connections between an EHR schema and FHIR are invalidated by updates to the FHIR specification. It is therefore desirable to automate this process, using only heuristic knowledge of each EHR schema. To this end, this paper introduces a semi-autonomous pipeline, which starts with a raw EHR, and produces a version of it that is compliant with the latest version of FHIR, according to a pre-defined set of parameters designed to encode partial knowledge about the schema of the EHR being converted. This pipeline is then realised as a \hl{piece of software, designed to be deployed into (decision-support) systems that use live EHR data.}

The remainder of this paper is organised as follows. We first review several background concepts, before introducing the theoretical components of our approach to identifying equivalent FHIR elements. We then detail how this approach is combined with both a pre-processing and post-processing (mapping) phase, in order to realise the system as a computational transformation pipeline. With the pipeline implemented, we then show how it can be applied and tuned to different EHR systems using different parameter settings.

\section{Background}\label{sec:background}

\subsection{EHR}

When structured as \textit{XML}\footnote{\emph{XML} stands for \emph{eXtensible Markup Language}, which is a widely employed methodology for describing and representing structured data sets. \hl{XML is versatile, and can be translated to and from a number of others standard formats, including JSON.}}, an EHR looks similar to the record shown in Figure \ref{fig:ehr}. We assume that any given fragment of XML from a vendor's EHR software refers to a single patient. Here, the names of tags and attributes and the relationships between elements (the \textit{data schema}), as well as the expected format of the data held in these elements, is dictated by the vendor's internal data model. Elements are either designed to hold additional, nested elements (e.g. the element with the tag name \textit{Demographics}), \hl{and are often referred to as \textit{complex types}}, or are designed to hold a piece of information relating to the patient (a \textit{leaf} element), such as the element with the tag name \textit{DateOfBirth}. For simplicity, we refer to a leaf element simply as an \textit{element}, an element holding nested elements as a \textit{parent element} and a tag as a \textit{name}. In addition, we refer to a single element, or a set of elements, identified by a name, by simply using that name, e.g. \textit{DateOfBirth}, to indicate the element with the name \textit{DateOfBirth} in Figure \ref{fig:ehr}. 

\begin{figure}[]
\centering
\begin{minipage}{.6\textwidth}
  \lstset{language=XML}
  \begin{lstlisting}[basicstyle=\small]
  <Record>
  <Demographics>
   <FirstName>Alice</FirstName>
   <Surname>Abbot</Surname>
   <DateOfBirth>1951-01-01</DateOfBirth>
   <Sex>F</Sex>
   <Address>
    <Text>1 Abbot Street</Text>
    <Country>United Kingdom</Country>
    <PostCode>ZZ99 3VZ</PostCode>
   </Address>
   <ManagingOrganisation>NHS</ManagingOrganisation>
  </Demographics>
  <Event>
    <Text>Patient first visit</Text>
    <Code>X3003</Code>
  </Event>
  <Event>
    <Code>X3003</Code>
  </Event>a
  </Record>
  \end{lstlisting}
  \end{minipage}
  \caption{Example EHR fragment}
  \label{fig:ehr}
\end{figure}

In an EHR XML fragment, parent elements with the same name may be repeated an arbitrary number of times, with potentially different child elements (e.g. the repetition of multiple \textit{Observation} elements seen in Figure \ref{fig:ehr}). Moreover, an element with the same name might appear multiple times with different parent elements (e.g. \textit{Text} appears as a child of both \textit{Address} and \textit{Observation}).

\subsection{FHIR}

Health Level 7, or \emph{HL7}\footnote{\url{http://www.hl7.org}}, is an international standards organisation that focuses on `exchange, integration, sharing, and retrieval of electronic health information'. HL7's \emph{Fast Healthcare Interoperability Resources (FHIR)} specification offers a lightweight alternative to their extensive HL7 v3 specification. Key medical entities are represented as individual \textit{resources}, specified in a manner that makes them easy to implement (e.g. as JSON schema specifications), with many example implementations made available as a part of the specification \cite{Bender2013}. The exchange of these resources, once populated, is conducted according to a standard RESTful\footnote{`REST' stands for \emph{REpresentational State Transfer}, which describes a methodology for data transfer using internet protocols~\cite{Fielding2000}} approach. Although there are limitations to the FHIR specification (for example, it is not currently clear how individual resources are combined in order to represent more complex medical concepts, something HL7's v3 defines using its \textit{Reference Information Model}), the \emph{agile}~\cite{Cockburn2002} software development process associated with FHIR is designed to enable incremental improvements to the specification.

\begin{figure}[]
\centering
\begin{tabular}{cc}
\begin{minipage}{0.6\textwidth}
  \lstset{language=XML}
  \begin{lstlisting}[basicstyle=\small]
<Patient>
 <id>001</id>
 <name>
  <given>Alice</given> [String]
  <family>Abbot</family>
 </name>
 <birthDate>1951-01-01</birthDate> [Date yyyy-mm-dd]
 <gender>female</gender> [male | female | other | unknown]
 <Address>
  <text>1 Abbot Street</text>
  <country>United Kingdom</country> [String]
  <postalCode>ZZ99 3VZ</postalCode>
 </Address>
 <managingOrganization>
    <identifier>002</identifier>
 </managingOrganization>
</Patient>
  \end{lstlisting}
\end{minipage}%
\end{tabular}
\caption{Example \textit{Patient} FHIR resource.}
\label{fig:fhir-p}
\end{figure}
\begin{figure}[]
\centering
\begin{tabular}{cc}
\begin{minipage}{.6\textwidth}
   \lstset{language=XML}
   \begin{lstlisting}[basicstyle=\small]
       <Organization>
        <id>002</id>
        <name>NHS</name>
       </Organization>
       
       <Observation>
        <subject>
         <identifier>001</identifier>
        </subject>
        <note>
         <text>Patient first visit</text>
        </note> 
        <code>
         <coding>X3003</coding>
        </code>
        <status>final</status>
       </Observation>
       
   \end{lstlisting}
\end{minipage}
\end{tabular}
\caption{Example \emph{Organization} and \hl{\mbox{{\textit{Observation}}}} FHIR resources.}
\label{fig:fhir-oe}
\end{figure}

\begin{figure}[]
\ContinuedFloat
\centering
\begin{tabular}{cc}
\begin{minipage}{.5\textwidth}
   \lstset{language=XML}
   \begin{lstlisting}[basicstyle=\small]

       <Observation>
        <subject>
         <identifier>001</identifier>
        </subject>
        <code>
         <coding>X3003</coding>
        </code>
        <status>final</status>
       </Observation>
       
   \end{lstlisting}
\end{minipage}
\end{tabular}
\caption{Example \emph{Organization} and \hl{\mbox{{\textit{Observation}}}} FHIR resources.}
\label{fig:fhir-oe}
\end{figure}

Like EHR data, each FHIR resource can be realised as an XML fragment, as shown in the four \textit{simplified} examples in Figures \ref{fig:fhir-p} and \ref{fig:fhir-oe} for \textit{Patient}\footnote{\url{https://www.hl7.org/fhir/patient.html}},  \textit{Organization}\footnote{\url{https://www.hl7.org/fhir/organization.html}} and \textit{Observation}\footnote{\url{https://www.hl7.org/fhir/observation.html}}. Here, parent elements are either resources (e.g. \textit{Patient}); \textit{datatype} resources, designed to exist as part of other resources (e.g. \textit{Address}); or \textit{references} to resources, where an \textit{identifier} element is used to supply the ID of another instantiated resource. For example, we observe that  \textit{Patient} (Figure \ref{fig:fhir-p}) is connected to \textit{Organization} (Figure \ref{fig:fhir-oe}) because the field \textit{managingOrganization} in \textit{Patient} is designed to hold the ID of resources of type \textit{Organization}. Connected resources can, themselves, be connected to additional resources, thus creating a connected graph in which resources are nodes, ID references are edges, and one resource can potentially be reached from another by following a \textit{path} of connections through intermediate resources. 

Elements themselves are designed to hold a piece of information relating to a resource (e.g. elements such as \textit{country}). Additional information is also provided by the standard on the type of data that can be stored within each element using a regular expression (regex)\footnote{Regular expressions are defined as sequences of characters that describe patterns which are used to identify character combinations in strings.}. \hl{While the codes used to populate a FHIR element may be drawn from a variety of terminologies, here we enforce the use of the SNOMED CT vocabulary\footnote{\url{http://snomed.info/sct}}}. 

\subsection{SMART--on--FHIR}

The SMART platform paradigm\footnote{`SMART', in the Health Informatics literature, stands for \emph{Substitutable Medical Applications, Reusable Technologies}} is designed as a response to a perceived inflexibility in traditional healthcare platforms, such as EHR software, which offer the same functionality irrespective of their use (e.g., in hospitals, GP surgeries, etc.) \cite{Mandl2012}. The SMART platform paradigm posits that all healthcare platforms should be encapsulated in \textit{SMART containers}, in order to expose a recognised interface. As such, healthcare \textit{applications} can be developed that communicate with this interface -- and thus the encapsulated platform -- in order to extend that platform's functionality, much in the same way that applications extend the functionality of a smartphone. For example, a GP surgery could purchase a SMART EHR system, and then choose those applications that best tailor the system for its needs. SMART--on--FHIR augments this approach by suggesting that the interface should adhere to FHIR's communication standard, and should add features such as authentication \cite{Mandel2016}.

\section{Methods}\label{sec:methods}

When translating an arbitrary EHR to FHIR, our aim is to find a set of FHIR elements that can directly accommodate the data from our EHR. For example, the set of elements shown in Figures \ref{fig:fhir-p} and \ref{fig:fhir-oe} can accommodate all of the EHR data from Figure \ref{fig:ehr}. This is because the set of FHIR elements shown is \textit{equivalent} to the set of elements in the EHR. A set of FHIR elements is equivalent to a set of EHR elements if, when used to markup data, they communicate the same information about the content of a record. We refer to a set of EHR elements and equivalent FHIR elements as an \textit{equivalence set}. A set of EHR and FHIR elements forms an equivalence set if the following conditions hold:

\begin{enumerate}

    \item \label{equiv:1} The name of each element in the EHR set \textit{matches} at least one of the element names in the FHIR set. For example, the element \textit{country} in the EHR in Figure \ref{fig:ehr} and the element \textit{country} in the FHIR resource in Figure \ref{fig:fhir-p} communicate to us that they contain the same information because their names match. The same is true for \textit{sex} and \textit{gender}\footnote{Sex and gender are still considered equivalent in many standards} and \textit{DateOfBirth} and \textit{birthDate}.

    \item \label{equiv:2} The type of every EHR element is \textit{compatible} with the type of the FHIR element to which it is matched. For example the element \textit{Firstname} (Figure \ref{fig:ehr}) and the element \textit{given} (Figure \ref{fig:fhir-p}) communicate to us that that they contain the same type of information not just because their names match, but also because the (inferred) \textit{string} type of \textit{Firstname} is compatible with the \textit{string} type of \textit{given}. The same is true of the \textit{DateOfBirth} and \textit{birthDate} match, as the type of data held is the same, despite being in a different \textit{format}.

	\item \label{equiv:3} Because EHR elements are inherently connected by appearing in the same EHR, if the set of FHIR elements is split over a number of different resources, those resources must also be connected in order to be considered equivalent. For example, the set of EHR elements \textit{DateOfBirth} and \textit{Code} (Figure \ref{fig:ehr}) and the set of FHIR elements \textit{birthDate} and \textit{coding} (Figure \ref{fig:fhir-oe}) communicate to us that they contain the same information, not just because they contain two compatible matches, but also because, like the EHR elements, the FHIR elements are connected: they are associated with the parent elements \textit{Patient} and \textit{Observation}, respectively, where the latter is connected to the former via the \textit{subject} element (Figure \ref{fig:ehr}).

\end{enumerate}

\subsection{Matching}\label{sec:matching}

In order to identify if constraint \ref{equiv:1} holds, we need to determine when a term from an EHR matches a term from the FHIR standard. In order to do this, we need to measure the \textit{similarity} between two terms and then see if this similarity exceeds a chosen \textit{threshold}, which is thus indicative of a match. In reality, there are a number of different measures for similarity, each of which provides important insight into the connection between two terms, and thus a number of potentially different thresholds. In order to extract a single decision about whether two terms match, we therefore have a choice about which similarity measures we select, how we combine them, and which values should be used for different thresholds.

In this section, we introduce three different similarity measures, prompted by several heuristics as to how one can find a matching FHIR term given an EHR term; introduce mechanisms to control the way in which these measures are combined in order to derive an overall similarity value; and introduce several different thresholds that can be set to control the sensitivity of each measure. What this results in is a set of parameters that can be used to \textit{configure} the mechanisms that determine when two terms match. Later, we will show that having a set of configurable similarity metrics as the basis for a piece of automated EHR-FHIR matching software is important, as EHR schema are designed in different ways, and this needs to be taking into account when matching elements from different schema to FHIR.

\subsubsection{Text (syntactic) similarity}\label{sec:text}

In many cases, a match for an EHR term in FHIR can be found by simply identifying a FHIR term with the same name, e.g. \textit{Country} and \textit{country}, or the same name with different spelling, such as \textit{organisation} and \textit{organization}. Our first individual measure of similarity is therefore \textit{text} similarity, for which we rely on \hl{normalised} \textit{Levenshtein distance}~\cite{Hall1980}, written as:

\begin{equation} 
\label{equ:lev}
{lev_{stringA, stringB}(|stringA|, |stringB|)} \mathcolourbox{}{= 1 - {\frac{distance}{|stringA| + |stringB|}}}
\end{equation}

This measure broadly returns the similarity between two strings based upon their constituent characters. For example, for \textit{stringA=Country} and \textit{stringB=country}, $lev$ would return 1. For this similarity measure we introduce the threshold $\tau_{t}$, such that if $lev >= \tau_{t}$, we say that two strings match. For example, for \textit{stringA} and \textit{stringB}, this would be the case if $\tau_{t} = 1$. \hl{We select (normalised) Levenshtein distance as the basis for this similarity metric as it is deemed to provide a good balance between both efficiency and flexibility \mbox{\cite{Christen2007}}}. 

\subsubsection{Morphological similarity}\label{sec:morph}

One might also find suitable counterparts for an EHR term in FHIR by aiming to identify a FHIR term that demonstrates a \textit{morphological} variance. For example, \textit{stringA=post} and \textit{stringB=postal} should be connected. \textit{lev} introduces an important level of potential flexibility, in that lower values of $\tau_{t}$ -- such as $\tau_{t} = 0.6$ in this instance -- could also be used to identify such a connection. However low values of $\tau_{t}$ are likely to lead to false positives for regular text similarity (e.g. when \textit{stringA=Country} and \textit{stringB=County}).

We therefore introduce a second similarity method for morphological similarity, detailed in Algorithm \ref{alg:a1}. Here, morphological variants of \textit{stringA} are generated and compared with \textit{stringB} using \hl{{\textsc{Lev}}, an algorithmic realisation of Equation {\ref{equ:lev}}. Several of these algorithms exist (e.g. \mbox{\cite{Paterson1980}})}. This connection with $lev$ represents the first example of two of our measures of similarity being combined. To generate morphological variants, the set of lemmas (a set containing a canonical word, repeated for different meanings) that are associated with \textit{stringA} are produced, followed by the set of lexemes (different word forms, including derived terms) associated with each lemma. For example, with an input of \textit{post}, a set of lemmas containing \textit{post} (mail) and \textit{post} (advertisement) would first be generated, followed by their related forms, such as \textit{postal} and \textit{posting}, respectively. \hl{Assuming the existence of a suitable mechanism for the generation of lemmas and lexemes,} the loops on lines \ref{alg:a1:loopA} and \ref{alg:a1:loopB} generate these variants, such that with the example input of \textit{stringA=post} and \textit{stringB=postal}, the conditional on line \ref{alg:a1:if} is eventually satisfied by a suitable variant. In this algorithm, we again see the use of the parameter $\tau_{t}$, which enables us to identify when the value returned by the text similarity metric is indicative of a string match. The output from this metric is derived from the morphological variant with the highest match.

\begin{algorithm}
\caption{morphologicalSimilarity}
\label{alg:a1}
\begin{algorithmic}[1]
\Require stringA, stringB, $\tau_{t}$
\If{\Call{lev}{stringA, stringB} $>=$ $\tau_{t}$} \State{return} 0 \EndIf
\State{highestSimilarity = 0}
\For{lemma of stringA} \label{alg:a1:loopA} 
\For{lexeme of lemma} \label{alg:a1:loopB} 
\If{\Call{lev}{lexeme, stringB} $>=$ $\tau_{t}$ \textbf{and} \Call{lev}{lexeme, stringB} $>$ highestSimilarity} \label{alg:a1:if} \State{highestSimilarity = \Call{lev}{lexeme, stringB}} \EndIf 
\EndFor
\EndFor
\State{return highestSimilarity}
\end{algorithmic}
\end{algorithm}

\subsubsection{Semantic similarity}\label{sec:semantic}

Another way in which one could find a counterpart for an EHR term in FHIR is to identify FHIR terms that \textit{mean} the same thing as the target EHR term. For example \textit{sex} and \textit{gender}. With this in mind, our third measure of similarity is \textit{semantic} similarity. This metric operates in a similar manner to morphological similarity, except rather than generating morphological variants of \textit{stringA} it instead \hl{assumes the existence of a suitable method} to generate the synonyms of \textit{stringA}. Producing a value for semantic similarity is thus a variant of Algorithm \ref{alg:a1}, shown in Algorithm \ref{alg:a2}. As we now have two similarity metrics, we have a choice as to which of them is used to identify when one of these synonyms matches \textit{stringB}. We choose \textsc{Lev} and its companion threshold value $\tau_{t}$ for this purpose, but we could have also used \textsc{\nameref{alg:a1}} and a new threshold value $\tau_{m}$. Using \textsc{Lev} as the matching metric, with an input of \textit{stringA=sex} and \textit{stringB=gender}, the loop shown on line \ref{alg:a2:loop} of Algorithm \ref{alg:a2} would eventually generate the word \textit{gender}, satisfying the condition on line \ref{alg:a2:if}. At the start of the algorithm (Line \ref{alg:a2:similar}) our text similarity metric also allows us to identify when the constituent characters of two words are already sufficiently similar (i.e., there is a text match), and thus do not need to be compared semantically, again illustrating the initial combination of different similarity metrics.

\begin{algorithm}{}
\caption{semanticSimilarity}
\label{alg:a2}
\begin{algorithmic}[1]
\Require stringA, stringB, $\tau_{t}$
\If{\Call{lev}{stringA, stringB} $>=$ $\tau_{t}$} \label{alg:a2:similar} \State{return 0} \EndIf  
\State{highestSimilarity = 0}
\ForAll{synonym of stringA} \label{alg:a2:loop} 
\If{\Call{lev}{synonym, stringB} $>=$ $\tau_{t}$ \textbf{and} \Call{lev}{synonym, stringB} $>$ highestSimilarity} \label{alg:a2:if} 
\State{highestSimilarity = \Call{lev}{synonym, stringB}} 
\EndIf 
\EndFor
\State{return highestSimilarity}
\end{algorithmic}
\end{algorithm}

\subsubsection{Compound words}\label{sec:compound}

While each of the metrics listed allows us to identify the existence of a connection between two strings, they are not designed to compare compound words. For example, while \textit{PostCode} and \textit{postalCode} are morphologically similar, they would not be reported as such by our morphological similarity metric, due to presence of compound words. Each of the algorithms can therefore be used in conjunction with Algorithm \ref{alg:a3} which identifies the similarity between compound words by breaking them down and then performs pairwise comparisons using one of the above metrics. For example, with \textit{stringA=PostCode} and \textit{stringB=postalCode}, both compound words would be separated, allowing \textit{post} to be connected to \textit{postal} using morphological similarity. The use of pairwise comparisons also allows us to deal with cases in which a FHIR term is a suitable candidate for an EHR term because the same compound words are used, but they are in a different order. For example when \textit{stringA=DateOfBirth} and \textit{stringB=birthDate}.

In our first instance of controlling how individual similarity metrics are combined, we introduce two mechanisms to interpret the overall similarity of two compound words: either the average similarity across all pairwise comparisons (Line \ref{alg:a3:average}), or the highest similarity across all pairwise comparisons (Line \ref{alg:a3:highest}). Which of these is chosen is controlled by the boolean parameter $\alpha$.

\begin{algorithm}
\caption{compositeStringSimilarity}
\label{alg:a3}
\begin{algorithmic}[1]
\Require stringA, stringB, \textsc{similarityMetric}, $\alpha$, \algcolor{\highlightcolour}{delimiter}
\State{highestSimilarity = 0}
\State{totalSimilarity = 0}
\ForAll{words in \algcolor{\highlightcolour}{\Call{tokenize}{stringA, delimiter}}}
\State{highestWordSimilarity = 0}
\ForAll{words in stringB}
\If{\Call{similarityMetric}{wordA, wordB} $>$ highestSimilarity}
\State{highestSimilarity=\Call{similarityMetric}{wordA, wordB}} \EndIf
\If{\Call{similarityMetric}{wordA, wordB} $>$ highestWordSimilarity} \State{highestWordSimilarity=\Call{similarityMetric}{wordA, wordB}} \EndIf
\EndFor
\State{totalSimilarity += highestWordSimilarity}
\EndFor
\If{$\alpha$}\State{return totalSimilarity / length(stringA)} \label{alg:a3:average} \Else \State{return highestSimilarity} \label{alg:a3:highest} \EndIf
\end{algorithmic}
\end{algorithm}

\subsubsection{Identifying a match}\label{sec:match}

In the previous sections we introduced three algorithms designed to provide different perspectives on the similarity between terms. Algorithm \ref{alg:a3} then gave us the capability to apply \textit{one} of our initial three individual similarity measures to terms with composite words. However, in many cases we would like to apply more than one similarity measure to the same composite terms. For example, the compound words in \textit{PostCode} and \textit{postalCode} are matched by morphological and text similarity, respectively, and we want to identify this.

Algorithm \ref{alg:a4} is designed for this purpose, and is used to determine when two terms match. Using Algorithm \ref{alg:a3}, it first computes the text, morphological and semantic similarity of two potentially composite terms. The way in which these individual metrics are combined is controlled by a series of additional parameters. First we introduce our remaining threshold, $\tau_{s}$, to identify when semantic values are indicative of a suitable match, as well as a main threshold, $\tau$, which is used irrespective of metric. Next, we introduce \textit{weightings} for each metric ($\omega_{t}, \omega_{m}, \omega_{s}$), in order to allow us to control the influence each has when determining if two names match.  Finally, the way in which our metrics are actually combined can be controlled using the remaining (boolean) parameters shown in Table \ref{table:params}.

\begin{algorithm}
\caption{match}
\label{alg:a4}
\begin{algorithmic}[1]
\Require stringA, stringB, $\tau$, $\tau_{t}$, $\tau_{m}$, $\tau_{s}$, $\omega_{t}$, $\omega_{s}$, $\omega_{g}$, $\beta$, $\gamma$, $\delta$, $\epsilon$.
\State{textSim = \Call{\nameref{alg:a3}}{stringA, stringB, lev} * $\omega_{t}$}
\If{ $\beta$ \textbf{and} textSim $>=$ $\tau_{t}$ } \State{return textSim }\EndIf
\State{morphologySim = \Call{\nameref{alg:a3}}{stringA, stringB, morphologicalSimilarity} * $\omega_{g}$}
\If{ $\beta$ \textbf{and} morphologySim $>=$ $\tau_{g}$ } \State{\Return{morphologySim} }\EndIf
\State{semanticSim = \Call{\nameref{alg:a3}}{stringA, stringB, semanticSimilarity} * $\omega_{s}$}
\If{ $\beta$ \textbf{and} semanticSim $>=$ $\tau_{s}$ } \State{\Return{semanticSim} }\EndIf
\If{ $\beta$ } \State{\Return{0}} \EndIf
\algstore{bkbreak}
\end{algorithmic}
\end{algorithm}

\begin{algorithm}
\ContinuedFloat
\caption{match (continued)}
\label{alg:a4}
\begin{algorithmic}[1]
\algrestore{bkbreak}
\State{maxSim=$\max$(textSim, semanticSim, morphologySim)}
\State{totalSim=textSim + semanticSim + morphologySim}
\State{averageSim=textSim + semanticSim + morphologySim / 3}
\If{ $\gamma$ \textbf{and} maxSim $>$ $\tau$ } \State{\Return{maxSim}}
\ElsIf{ $\delta$ \textbf{and} totalSim $>$ $\tau$ } \State{\Return{totalSim}}
\ElsIf { $\epsilon$ \textbf{and} averageSim $>$ $\tau$ } \State{\Return{averageSim}} \EndIf
\State{\Return{0}}
\end{algorithmic}
\end{algorithm}

\begin{table}[ht!]

	\begin{center}

		\begin{tabular}{p{.15\columnwidth}p{.1\columnwidth}p{.65\columnwidth}}
			\hline
            Parameter & Type & Description \\
            \hline
			$\tau_{t}$ & float & Text match threshold (\textsc{Lev})  \\
            $\tau_{m}$ & float & Morphological match threshold (\textsc{\nameref{alg:a1}}) \\
            $\tau_{s}$ & float & Semantic match threshold (\textsc{\nameref{alg:a2}}) \\
            $\omega_{t}$ & float & Text similarity weighting \\  
            $\omega_{m}$ & float & Morphological similarity weighting \\ 
            $\omega_{s}$ & float & Semantic similarity weighting \\
            $\alpha$ & boolean & Average similarity or highest similarity, when calling \textsc{\nameref{alg:a3}}. \\
            $\beta$ & boolean & Do not combine metrics, but instead use the first metric that exceeds its threshold (assuming \textsc{Lev} $>$ \textsc{\nameref{alg:a1}} $>$ \textsc{\nameref{alg:a2}} in terms of importance). \\
            $\gamma$ & boolean & When combining metrics,  use the metric with the highest strength.  \\
            $\delta$ & boolean & Combine the results of all of the metrics.  \\
            $\epsilon$ & boolean & Return the average result. \\
            \hline
		\end{tabular}
		\caption{An overview of the available parameters in our system.}
		\label{table:params}

	\end{center}

\end{table}%

\subsection{Data type compatibility}\label{sec:datatype}

We now have an algorithm to determine whether an EHR term and a FHIR term match, thus enabling us to determine our first aspect of equivalence, listed above. The second way in which two terms can be considered equivalent is if they store compatible types of data. In order to identify if the type of a FHIR element is compatible with the data stored in an EHR element, we rely on the specification of FHIR types as regular expressions from the FHIR standard. If the data in an EHR element matches the regex associated with the tag used by a FHIR element, the EHR data can be legally stored within an element using that tag. However, the specificity of FHIR datatype regexes means that a piece of EHR data might fail a regex, despite still being of a type that can be \textit{reformatted} in order to match the format of the expression. For example, the date regex provided by the FHIR standard enforces the pattern \textit{yyyy-mm-dd}, which would exclude the date \textit{dd-mm-yyyy}, despite it being possible to reformat data expressed in this way in order to comply with the FHIR standard. We therefore introduce a secondary set of more general regular expressions that are able to identify data that is of a type and format that can be reformatted in order to comply with the type of a given FHIR element when mapping. An example of some these expressions is shown in Table \ref{fig:reg}. Therefore, when determining the type compatibility of a match, a table like the one shown is first indexed using the matched FHIR element's data type, the general expression is extracted and applied to the content of the source EHR element, and if it passes it is considered compatible, otherwise it is considered incompatible.

\begin{table}[ht!]

	\begin{center}

		\begin{tabular}{p{.1\columnwidth}p{.4\columnwidth}p{.4\columnwidth}}
			\hline
            Element & Format regex & General format regex \\
            \hline
			Date & $-?[0-9]{4}(-(0[1-9]|1[0-2])(-(0[0-9]|[1-2][0-9]|3[0-1]))?)?$ & $(?=.{10})([0-9]{4}|[1-2][0-9]|3[0-1])(-(0[1-9]|1[0-2])(-(0[0-9]|[1-2][0-9]|3[0-1]|[0-9]{4}))?)?$ \\
            Gender (Code) & $(male|female|other|unknown)$ & $([Mm]+[A-Za-z]*|[Ff]+[A-Za-z]*|[Oo]+[A-Za-z]*|[Uu]+[A-Za-z]*)$ \\
            \hline
		\end{tabular}
		\caption{Example regular expressions to identify EHR data that can be transformed to comply with a FHIR element.}
		\label{fig:reg}

	\end{center}

\end{table}%

\subsection{Identifying an equivalence set}\label{sec:set}

We now have the means to identify if an FHIR term is able to communicate the same information as an EHR term based upon two of the three equivalence constraints listed. To satisfy the third, we require a process that derives a configuration of compatible EHR to FHIR element matches, while ensuring that they are all contained within connected resources. With all three constraints satisfied, we consider a configuration a valid equivalence set. The identification of an equivalence set is presented in Algorithm \ref{alg:a5}. The aim is to find the largest number of compatible FHIR element matches for the set of EHR elements that exist within connected resources, thus satisfying all equivalence constraints. To do this, the algorithm examines each FHIR resource in turn, and then, for each EHR element, determines if the elements of this resource (Line \ref{alg:a5:thisclass}) or any of its connected resources (Line \ref{alg:a5:otherclasses}) contain a compatible match. If multiple valid matches happen to be found for an individual EHR element, one is selected at random. After this process is complete, we are left with a number of equivalence sets, each containing a set of connected compatible matches. The set with the highest number of matches is returned from the algorithm. If multiple sets contain the same number of matches, the set to return is determined based upon on the smallest distance between selected resources and then upon the fewest number of overall resources used to house the EHR elements, both of which suggest a more intelligible representation.

\begin{algorithm}[]
\caption{getEquivalenceSet}
\label{alg:a5}
\begin{algorithmic}[1]
\Require ehrElements, fhirElements, fhirResources, fhirConnections
\ForAll{fhirResources} 
\ForAll{ehrElements, ehrParents} 
\State{ set $\gets$ \Call{getCompatibleMatch}{fhirResource, fhirElements, fhirConnections, ehrElement} }
\EndFor
\State{allSets $\gets$ set}
\EndFor
\State{\Return sort(allSets)}
\Statex
\Function{getCompatibleMatch}{fhirResource, fhirElements, fhirConnections, ehrElement} \funclabel{getCompatibleMatch}
    \ForAll{fhirElements[fhirResource]}\label{alg:a5:thisclass}
    	\If {\Call{\nameref{alg:a4}}{fhirElement, ehrElement} $>$ 0 \textbf{and} \Call{dataTypeCompatible}{fhirElement, ehrElement}} 
        	\State{\Return fhirElement:ehrElement}
        \EndIf
    \EndFor
    \ForAll{fhirConnections}\label{alg:a5:otherclasses}
    	\State{\Call{getCompatibleMatch}{connectedFHIRResource, fhirElements, fhirConnections, ehrElement}}
    \EndFor
    \State{\Return None}
\EndFunction
\end{algorithmic}
\end{algorithm}

\subsection{Transformation Pipeline}\label{sec:transformation}

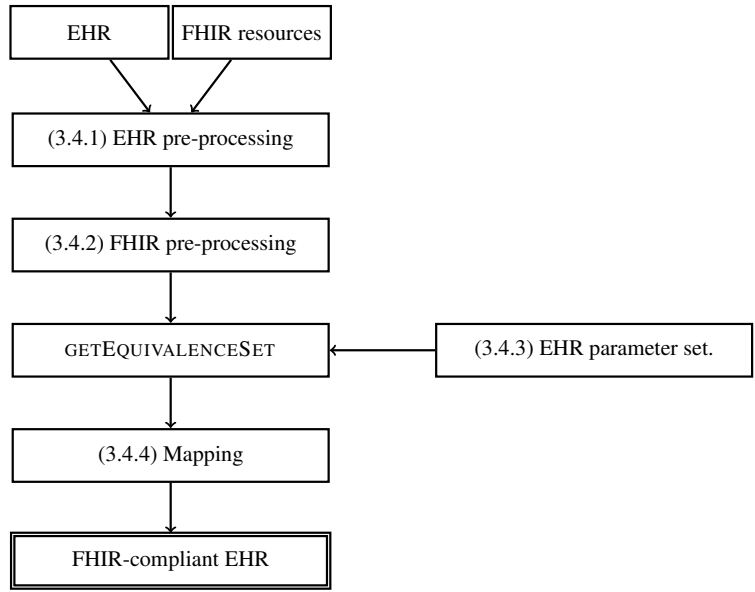
\begin{figure}[ht!]

\tikzset{rectangle state/.style={draw,rectangle}}

\begin{center}

\centering\scalebox{0.7}{

\begin{tikzpicture}[node distance=2cm,initial text=]

\tikzstyle{every node}=[font=\large]

\begin{scope}

\node[rectangle state,very thick,minimum width=60mm,minimum height=10mm, align=left] (ehr-pre)  {(\ref{sec:ehr-preprocessing}) EHR pre-processing};
\node[rectangle state,very thick,minimum width=30mm,minimum height=10mm] (ehr)[above left = 1cm and -3cm of ehr-pre]  {EHR};
\node[rectangle state,very thick,minimum width=30mm,minimum height=10mm] (fhir) [above right = 1cm and -3cm of ehr-pre] {FHIR resources};
\node[rectangle state,very thick,minimum width=60mm, align=left,minimum height=10mm] (fhir-pre) [below of=ehr-pre] {(\ref{sec:fhir-preprocessing}) FHIR pre-processing};
\node[rectangle state,very thick,minimum width=60mm, align=left,minimum height=10mm] (alg) [below of=fhir-pre] {\textsc{\nameref{alg:a5}}};
\node[rectangle state,very thick,minimum width=60mm,minimum height=10mm] (param) [right=2cm of alg] {(\ref{sec:configuration}) EHR parameter set.};
\node[rectangle state,very thick,minimum width=60mm, align=left,minimum height=10mm] (mapping) [below of=alg] {(\ref{sec:mapping}) Mapping};
\node[rectangle state,accepting,very thick,minimum width=60mm, align=left,minimum height=10mm] (complete) [below of=mapping] {FHIR-compliant EHR};

\end{scope}

\begin{scope}

\path[->] (ehr) edge [very thick] node [] {} (ehr-pre);
\path[->] (fhir) edge [very thick] node [] {} (ehr-pre);
\path[->] (ehr-pre) edge [very thick] node [] {} (fhir-pre);
\path[->] (fhir-pre) edge [very thick] node [] {} (alg);
\path[->] (param) edge [very thick] node [] {} (alg);
\path[->] (alg) edge [very thick] node [] {} (mapping);
\path[->] (mapping) edge [very thick] node [] {} (complete);

\end{scope}

\end{tikzpicture}}

\caption{An overview of the EHR transformation process.}

\label{fig:pipe}

\end{center}

\end{figure}

In order to fully define the process for converting a patient record to FHIR, we combine \textsc{\nameref{alg:a5}} with several other processing stages in order to form a \textit{transformation pipeline}, shown in Figure \ref{fig:pipe} and described below.

\subsubsection{EHR pre-processing}\label{sec:ehr-preprocessing}

First, elements from the EHR are aggregated in order to be matched. \hl{During this process, elements and parent elements with \textit{coded} name values are identified. Naturally, coded name values are not be amenable to our matching approach. Therefore, we rely on external services to provide us with additional information on any coded name values, specifically their associated text representation (e.g. SNOMED code \textit{184099003} to \textit{DateOfBirth}). This information is then used to convert any coded names values to their text equivalent, so that they can also be processed by the rest of our pipeline. Examples of these services include UMLS~\mbox{\cite{Bodenreider2004}}, BioPortal~\mbox{\cite{Whetzel2011}} or MetMaps~\mbox{\cite{Liang2016}}.} 

If an element with the same name appears multiple times under the same parent (e.g. \textit{Code} under \textit{Observation} in Figure \ref{fig:ehr}) it is only listed once. Elements with the same name that are listed under different parents are deemed to be too general to sufficiently communicate information about the data they contain. For example it is not clear, simply from the element \textit{Text}, whether this is being used in the EHR to markup the text of an address or the text of an event description. Therefore, these elements are altered to also include their parent name, in this case to produce the new elements \textit{AddressText} and \textit{ObservationText}. This approach ensures that elements like this are more likely to be matched with FHIR elements that correctly relay their content.

\subsubsection{FHIR pre-processing}\label{sec:fhir-preprocessing}

Next, each FHIR resource is linked to a list of its elements in order to be used in the construction of an equivalence set. During this step, parent elements that contain references to other resources are also included in the list of elements for a resource if they are \textit{effectively leaf nodes}. That is, matching an EHR element to the name of a FHIR resource link is not usually useful, as information cannot be stored in a resource with that name directly, but instead only within one of its elements. For example, matching the EHR element \textit{ManagingOrganisation} to the field \textit{managingOrganization} in FHIR does not necessarily help us, as \textit{managingOrganization} is simply a reference to an \textit{Organization} resource, not a place in which data can be stored directly. However, if a target resource has an element with a name that indicates it can store general data about the resource, such as \textit{name}, which classes like Organization do, then matching with the parent is sufficient, as data can be held in this child, effectively making the parent a leaf element. Therefore, we aggregate both the elements of each resource, and their effective leaf nodes.

Much like EHR pre-processing, our next FHIR pre-processing step involves selectively augmenting the names of FHIR elements with their parent name in order to better dictate their context. We see this issue again in Figure \ref{fig:fhir-p}, where the element with the tag name \textit{given} does not communicate alone the same information as it does when coupled with its parent \textit{name}: that in reality this tag means \textit{given name}. However, in response to this issue, to ensure we are only adding to the specification, rather than altering the names of such elements to include their parent name directly, instead additional \textit{pseudo} elements are added to the list of FHIR elements, with names that are a combination of child and parent names (\textit{givenName}).

Finally, as a part of our pre-processing phase for FHIR resources, we establish which links exist between our set of resources, in order to later ascertain if constraint \ref{equiv:3} holds.

\subsubsection{Parameter configuration}\label{sec:configuration}

Our third input to the pipeline is a mapping between each piece of EHR software that will interact with the pipeline and a parameter configuration. The available parameters are summarised in Table \ref{table:params}. Depending on which piece of EHR software the record being input to the pipeline originates from, a different set of parameters will be used for the matching process. In this way, the matching process can be tuned to the perceived idiosyncrasies of the inferred EHR schema. For example, the administrator of a SMART--on--FHIR container might observe that the records generated from the encapsulated EHR tend to use terms that are semantically related yet identify different data. During our own evaluation, we observed this phenomenon in SystmOne records where, for example, the tags \textit{Address} and \textit{Reference} are commonly used to identify different data, despite being semantically related (in this case via a computing context). This observation can be encoded by decreasing the influence of the semantic similarity metric ($\omega_{s}$) on the overall match score when converting SystmOne records.

\subsubsection{Mapping}\label{sec:mapping}

Once an equivalence set is generated, an empty set of XML documents is produced, representing the elements of the equivalent FHIR elements and their parent resources, ready to receive data extracted from the EHR. Extraction then involves the following stages:

\begin{enumerate}
    
	\item Identify any data contained in matched EHR elements that is of the correct type, but requires transformation to the correct format, using a general format regex, such as one of the ones shown in Table \ref{table:transform}. Apply the transformation associated with the matching regex to the data, in order to convert it to the format required by FHIR.
    
    \item Transform any coded data values that are not compatible with FHIR (typically data that is not encoded using SNOMED CT) is transformed using a vocabulary service mentioned in Section \ref{sec:ehr-preprocessing}.

    \item Map data contained in EHR elements that have been connected to pseudo-elements to their real counterparts, for example if \textit{FirstName} is connected to \textit{givenName}, the data within \textit{FirstName} will be mapped to the \textit{given} FHIR element, within \textit{name}.

    \item Map data contained in EHR elements that has been connected to effective leaf nodes to the actual element within that resource.

    \item Combine the data contained in two or more EHR elements that have been connected to the same FHIR element and map the combined data to that element.
    
    \item \hl{Identify any missing mandatory elements in each resource and supply a default value. For example, if the mandatory \textit{status} element in the \textit{Observation} resource were not the target of a match, but other fields were, then a default value would be used, such as \textit{final}. This can be see in Figure \mbox{\ref{fig:fhir-oe}}}.

    \item Construct reference numbers and reference fields (Figures \ref{fig:fhir-p} and \ref{fig:fhir-oe}), for each connected resource and connecting resource, respectively.

\end{enumerate}

\begin{table}[ht!]

	\begin{center}

		\begin{tabular}{p{.1\columnwidth}p{.8\columnwidth}}
			\hline
            Type & Transformation \\
            \hline
			Date & Extract $[0-9]{4}$ (year), then $(0[1-9]|1[0-2])$ (month) and $(0[0-9]|[1-2][0-9]|3[0-1])$ date from the unformatted string, and then arrange them in the correct order, separated by hyphens.  \\
            Gender (Code) & Extract the first character of the EHR data (e.g. \textit{F}) and compare it to the first character of each of the options, selecting the one that matches (e.g. \textit{F} in \textit{female}). \\
            \hline
		\end{tabular}
		\caption{Example transformations based on matched FHIR format regexes.}
		\label{table:transform}

	\end{center}

\end{table}%

With elements and their elements now mapped to FHIR resources, and data extracted and transformed, the translation process is complete.

\subsection{Implementation}\label{sec:implementation}

In order to realise our transformation pipeline as a runnable system, we implement it as a piece of Python software. Our implementation is available under Apache 2.0 open-source licence \footnote{\url{https://github.com/consult-kcl/fhir-ehr-adapter}}. To implement our metrics, beyond Python's common constructs, we rely on the \textit{fuzzywuzzy} library\footnote{\url{https://github.com/seatgeek/fuzzywuzzy}} for text comparisons, and NLTK's\footnote{\url{https://www.nltk.org/}} \hl{\mbox{\textit{Wordnet}} interface \mbox{\cite{Miller1995}}}, for generating both morphological variants \hl{(lemmas and lexemes)} and synonyms. \hl{With our metrics implemented, we are able to computationally verify their accuracy using a broad range of test cases designed to reflect the types of data that will need to be matched between an EHR and FHIR. Under these test cases, our metrics are able to operate with $\geq$90\% accuracy.}

All of our interactions with the FHIR standard are via the \textit{fhir-parser}\footnote{\url{https://github.com/smart-on-fhir/fhir-parser}} library, which transforms the live FHIR specification into a set of Python classes. These classes are then parsed for use in our system. Therefore, matches are always based on the latest version of the FHIR specification.

\begin{figure}[]
\begin{center}
\centering\scalebox{0.7}{
\tikzset{rectangle state/.style={draw,rectangle}}
\begin{tikzpicture}[node distance=3.5cm,initial text=,
    database/.style={
      cylinder,
      shape border rotate=90,
      aspect=0.25,
      draw
    }]
\begin{scope}
\node[rectangle state,very thick,minimum size=10mm, align=center] (pipeline) {Transformation \\ pipeline};
\node[rectangle state,very thick,minimum size=10mm, align=center] (so-module) [above left of=pipeline]{SystmOne \\ module};
\node[rectangle state,very thick,minimum size=10mm, align=center] (so-installation) [left of=so-module]  {SystmOne \\ installation};
\node[database,very thick,minimum size=10mm, align=center] (so-data) [left of=so-installation] {SystmOne \\ formatted \\ data};
\node[rectangle state,very thick,minimum size=10mm, align=center] (m-module) [below left of=pipeline] {mMedica \\ module};
\node[rectangle state,very thick,minimum size=10mm, align=center] (m-installation) [left of=m-module] {mMedica \\ installation};
\node[database,very thick,minimum size=10mm, align=center] (m-data) [left of=m-installation] {mMedica \\ formatted \\ data};
\node[rectangle state,very thick,minimum size=10mm, align=center] (proxy) [right of=pipeline] {FHIR proxy};
\node[minimum size=10mm, align=center] (fhir-blank) [below of=proxy] {};
\node[database,very thick,minimum size=10mm, align=center] (fhir) [below= 0cm of fhir-blank] {FHIR \\ server};
\node[database,very thick,minimum size=10mm, align=center] (equivalence-set-cache) [left= 1.6cm of fhir] {Equiv. set \\ cache};
\node[rectangle,very thick,minimum size=10mm, align=center] (consult) [right of=proxy] {\consult \\ system};
\node[draw,dashed,inner sep=2mm,fit=(so-module)(m-module)(proxy),label={[above]HTTP Server}] {};
\node[draw,dashed,inner sep=2mm,fit=(so-data)(fhir)(proxy),label={[above]FHIR container}] {};
\end{scope}
\begin{scope}
\path[<->] (so-module) edge [very thick] node [below] {Tunnel} (so-installation);
\path[<->] (m-module) edge [very thick] node [above] {Tunnel} (m-installation);
\path[<-] (pipeline) edge [very thick] node [] {} (so-module);
\path[<-] (pipeline) edge [very thick] node [] {} (m-module);
\path[->] (so-installation) edge [very thick] node [] {} (so-data);
\path[->] (m-installation) edge [very thick] node [] {} (m-data);
\path[<-] (proxy) edge [very thick] node [] {} (pipeline);
\path[<-] (consult) edge [very thick] node [] {} (proxy);
\path[->] (pipeline) edge [very thick] node [] {} (fhir);
\path[<->] (pipeline) edge [very thick] node [] {} (equivalence-set-cache);
\path[->] (consult) edge [very thick] node [] {} (proxy);
\path[->] (proxy) edge [very thick] node [] {} (so-module);
\path[->] (proxy) edge [very thick] node [] {} (m-module);
\path[<->] (fhir) edge [very thick] node [] {} (proxy);
\end{scope}
\end{tikzpicture}}

\caption{A prototype of the \consult architecture using our software.}
\label{fig:consult-post}
\end{center}
\end{figure}
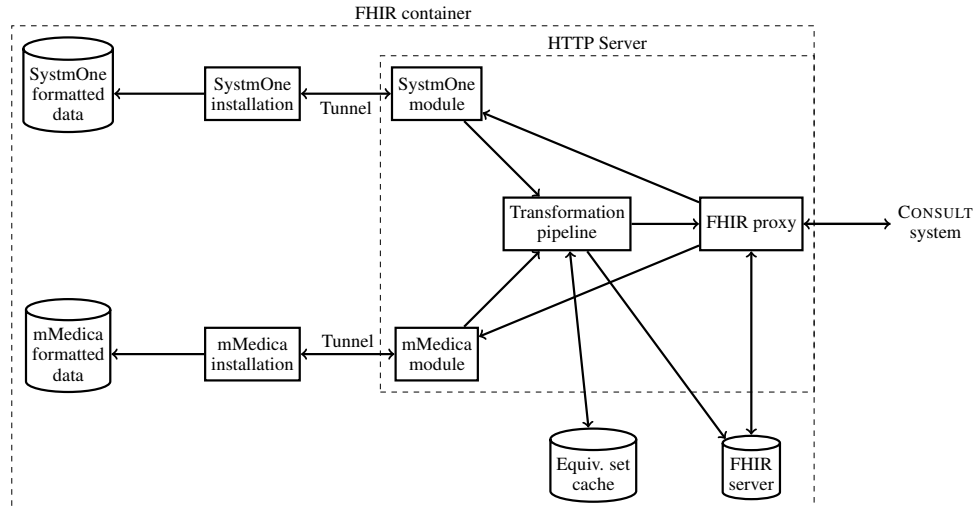

Recall that the \consult decision support system typifies the issue of interacting with the different standards used by different vendors. Figure \ref{fig:consult-post} shows how our implemented pipeline can be deployed into the \consult system as part of a SMART--on--FHIR container. Here, the rest of the system now communicates with a single FHIR endpoint for EHR data. To enable FHIR data to be served by the endpoint, our pipeline sits within a HTTP server, accessing the EHR software APIs on one side using custom modules supplied for each EHR, and proving an interface to the container on the other, using a proxy to a FHIR server (e.g. a HAPI-FHIR server\footnote{\url{http://hapifhir.io/}}). When a request for a new patient record is sent to the SMART--on--FHIR container, the FHIR proxy determines which piece of EHR software to contact for the record. Once it retrieves the record back from the EHR software, it passes it to the transformation pipeline. After identifying which elements from the EHR are to be matched, the pipeline first checks an equivalence set cache to see if this set of EHR elements have previously been connected to a set of FHIR elements. If they have, this set is used to perform the mapping process so that the computation required to identify these connections need not be repeated. Otherwise, the transformation pipeline runs as normal. Note that we do not cache connections individually, as a given EHR extract may not fully reflect the naming and structuring approach present in the rest of the schema.

Once the record has passed through the transformation pipeline, it is then stored in the FHIR server in order to be served by the FHIR proxy. Therefore, if the same record is requested in the future, it is served directly by the proxy from the cached copy in the FHIR server. This again reduces the computational load associated with the process of deriving an equivalence set, and with the mapping processes. Should any of the parameter sets be changed for a piece of EHR software, both the equivalence set cache and the FHIR server cache are emptied. Similarly, both caches are emptied if an update to the FHIR specification is detected. 

\section{Results and Discussion}\label{sec:application}

Our implemented transformation pipeline is able to operate \textit{semi-autonomously}, using the parameter set defined by the SMART--on--FHIR container administrator to convert any future records received back from that EHR to FHIR. We illustrate the operation of our transformation pipeline \hl{on a more complex data instance}, and the impact of different parameters sets, by applying our pipeline to a set of EHR records, taken from the \hl{public API} of two vendors available to the \consult system: SystmOne and mMedica. Given that FHIR covers a wide range of clinical entities, we select a subset of 60 classes that are most relevant to the representation of patient data. In addition, our selected EHRs contain 55 and 21 distinct elements for SystmOne and mMedica, respectively, and relate to patient demographics, along with information on their visits to the GP, recording, among other things, identified conditions, and the dispense of medication. \hl{An example of one of these EHRs can be found in \mbox{\ref{app:ehr}}}.

For both EHRs, we adopt the high value of .95 for each of our threshold parameters ($\tau_{t}$, $\tau_{m}$, $\tau_{s}$, and $\tau$) in order to prioritise accuracy when examining the impact of different configurations of our remaining parameters, but the administrator of a SMART--on--FHIR container who wishes, for example, for a higher number of matches with lower accuracy might lower these thresholds. Tables \ref{table:results-so} and \ref{table:results-m} show the portion of EHR elements that have been matched to FHIR elements, and related statistics, for each EHR, under a given parameter set\hl{, while Table {\ref{table:resources}} lists the range of resources to which the FHIR elements selected under all configurations belong}. In this set of parameter tests, we isolate our metric combination parameters by keeping our weights, and $\alpha$, constant. We observe that, for SystmOne, the impact of $\beta, \gamma$ and $\delta$ has a relatively small impact on the number of valid matches found, albeit $\delta$ offers the highest number of matches, at the expense of using slightly more resources. Thus, whichever the administrator selects as part of their configuration for SystmOne is of relatively little consequence. In contrast to SystmOne, for mMedica, it is clear that $\delta$ is the preferable option. For both of these EHRs, $\epsilon$ should be avoided. \hl{In both cases, the elements selected belong to resources ranging from those already seen, such as \textit{Patient} and \textit{Observation}, to other resources including \textit{Condition}, \textit{MedicationDispense} and \textit{DosageInstruction}.} Overall, we note that there is no perfect match for all of our EHR elements. \hl{For example, we observe that administrative elements of an EHR, such as version information and communication consent information, often fail to map to FHIR.} This is to be expected given the fact that FHIR is still evolving as a standard that is suitable for the representation of a wide range of EHR data.

\begin{table}[ht!]

	\begin{center}
		\centering
		\begin{tabular}{cccccccccccc}
			\hline
            $\omega_{t}$ & $\omega_{m}$ & $\omega_{s}$ & $\alpha$ & $\beta$ & $\gamma$ & $\delta$ & $\epsilon$ & \% matched & Resources & Distance \\
            \hline
            1 & 1 & .95 & T & T & F & F & F & .81 & 8 & 14.62  \\
            1 & 1 & .95 & T & F & T & F & F & .81 & 9 & 14.62 \\
            1 & 1 & .95 & T & F & F & T & F & .98 & 12 & 14.46 \\
            1 & 1 & .95 & T & F & F & F & T & .01 & 1 & 13 \\
            \hline
		\end{tabular}
		\caption{The impact of different parameter values on the processing of SystmOne records by our pipeline.}
		\label{table:results-so}

	\end{center}

\end{table}%

\begin{table}[ht!]

	\begin{center}
		\centering
		\begin{tabular}{cccccccccccc}
			\hline
            $\omega_{t}$ & $\omega_{m}$ & $\omega_{s}$ & $\alpha$ & $\beta$ & $\gamma$ & $\delta$ & $\epsilon$ & \% matched & Resources & Distance  \\
            \hline
            1 & 1 & 1 & T & T & F & F & F & .38 & 7 & 11 \\
            1 & 1 & 1 & T & F & T & F & F & .38 & 7 & 12.62 \\
            1 & 1 & 1 & T & F & F & T & F & .85 & 5 & 14.83 \\
            1 & 1 & 1 & T & F & F & F & T & 0 & - & - \\
            \hline
		\end{tabular}
		\caption{The impact of different parameter values on the processing of mMedica records by our pipeline.}
		\label{table:results-m}

	\end{center}

\end{table}%

\begin{table}[ht!]

	\begin{center}
		\centering
		\begin{tabular}{cccc}
			\hline
            Types & Individuals & Entities & Workflow  \\
            \hline
            HumanName & Patient & Organization & Task \\
            Address &  &  &   \\
            \cline{2-4}
            Attachment & Summary & Diagnostics & Medications \\
            \cline{2-4}
            Quantity & Observation & Condition & Medication  \\
            Period & & & MedicationDispense \\
            Identifier & & & DosageInstruction \\
            \hline
		\end{tabular}
		\caption{\hl{The types of FHIR resources selected by our pipeline, organised by category.}}
		\label{table:resources}

	\end{center}

\end{table}%

The process of investigating the impact of different parameter values shown in Tables \ref{table:results-so} and \ref{table:results-m} reflects the process that an administrator might go through in order to derive the parameter sets that are best for each piece of their EHR software. Although the parameters available are designed to allow an administrator to encode heuristic insight into each EHR schema, such as using $\omega_{s}$ to encode the semantic term phenomenon found in SystmOne, engaging in an experimental process to derive a set of parameters is an equally suitable way to derive a sufficient match rate for a given EHR. Even if parameters need to be refined several times in order to obtain suitable results, such an approach still offers an improvement over manually mapping every EHR (schema) element to FHIR, as only a small number of parameters need to be set and the the majority of the work is still carried out autonomously by the pipeline.

\subsection{Related Work}\label{sec:related}

The challenge of autonomously finding matches for each EHR element in FHIR can be framed as one of \textit{automated schema matching}, which has received significant attention in the research community, and a number of different mechanisms to identify connections between schema elements have been proposed \cite{Bernstein2011, Shvaiko2005}. However, the majority of holistic systems incorporating these mechanisms focus on \textit{ontology matching}, with only a small subset examining XML schema matching\footnote{We do not consider it to be trivial to effectively translate an (EHR) XML schema to an ontology.}. While popular software like \textit{COMMA++} \cite{Aumueller2005} does provide a means to compare XML schema, its focus is on how to combine existing matching mechanisms rather than the provision of those mechanisms themselves, and thus the sophistication of the matching mechanisms provided by this system, such as dealing with word morphology, is somewhat limited. Similarly \textit{Cupid} \cite{Madhavan2001}, although designed to deal with XML schema, focuses less on semantic similarity, while \textit{SASMINT} \cite{Madhavan2001} provides a detailed set of matching mechanisms, but does not offer the variety of weights used by our system to customise the matching mechanisms. In addition, the majority of these systems also lack the flexibility to exist in a multi-stage analysis pipeline like the one proposed, which includes operating on instances of data marked up using a schema, rather than the schema itself.

XML schema matching techniques have not previously been selected and combined in order to solve the problem of matching EHR extracts with FHIR. However, several other studies have examined the problem of converting health data (models) to FHIR to, for example, support the operation of a SMART--on--FHIR container. In each case, the matching process is conducted \textit{manually}. For example, Leroux et. al match ODM study data structures to FHIR \cite{Leroux2015}, in parallel with other initiatives such as \textit{OMOP--ON--FHIR}\footnote{\url{http://omoponfhir.org}}. Similarly, Duke Health encapsulate their EHR in a SoF container, matching their data elements to FHIR resources \cite{Bloomfield2017}, while Wagholikar et al. look at providing a FHIR interface to i2b2 repositories, expressing matches between data elements in these repositories and FHIR using \textit{XQuery} \cite{Wagholikar2017}. \hl{In addition, several software tools exist that aim to augment both the manual matching and mapping processes by providing a graphical interface through which a user can connect arbitrary data sources to FHIR. \textit{LinkEHR} (Studio) \mbox{\cite{Maldonado2009}} provides such an interface, as well as the ability to generate an XQuery transformation program that can be used for future conversions.}

\section{Conclusions and Future Work}\label{sec:future}

This work identifies a producer-consumer dichotomy in which producers, such as EHR systems, expose access to data formatted under complex internal models, while consumers, such as a decision support systems, require access to data formatted under lightweight messaging models, such as FHIR. We posit that while the use of FHIR containers is the most prudent way to connect these two models, there are limitations in the manual approach currently used within these containers to match elements of different EHR standards to elements in FHIR: scalability, FHIR schema variability, and the requirement for complete schema knowledge. To address the issue of scalability, we introduce a semi-automated matching and mapping pipeline, which is designed to convert EHR extracts directly to FHIR. To address insufficient schema knowledge, this pipeline does not require knowledge of the entities to be matched in advance, but instead only readily-available heuristic insight on the schema, expressed using a set of high-level parameters. Finally to address FHIR schema variability, our computational system processes live copies of the FHIR specification for use in the pipeline. To illustrate this approach, we use an implementation of our pipeline to transform a set of EHRs to FHIR, demonstrating the process that an administrator of the system might go through in order to derive a set of parameters that give rise to a sufficient match percentage for each of their chosen EHRs.

Not only does our approach contribute towards the operation of technology such as SMART--on--FHIR, it also makes a general contribution towards the movement to unify EHR standards, by potentially providing EHR companies with a means to gain insight into how they can connect their own schemas to FHIR. Moreover, by aiming to provide EHR data in a standardised format, our system increases the general accessibility of health data, for purposes such as research.

Future work will investigate if examining matches from across different EHRs within the pipeline can improve the quality of future matches. For example, if there is a relationship between two elements in different EHRs, then information about the connection of one of these elements to FHIR might be useful for connecting the other. From a usability perspective, future work will also aim to investigate the impact of involving the administrator with the translation pipeline more, such as to interactively review matches via a bespoke desktop tool \hl{or by integrating the interface to our pipeline with an existing non-automatic mapping tool, which would also complement its operation}. The involvement of a user becomes particularly important when a suitable match for a child cannot be found, and to provide feedback so that our system can \textit{learn} the optimal parameterisations for each piece of EHR software, such as the weightings given to each similarity metric.

\section{Acknowledgements}\label{sec:acknowledgements}

The authors are supported by the \consult project, funded by the UK Engineering \& Physical Sciences Research Council (EPSRC) under grant \#EP/P010105/1. The funding source had no involvement in this research.

\appendix

\section{Example source EHR}
\label{app:ehr}
\lstset{language=XML}
\begin{lstlisting}[basicstyle=\small]
<Identity>
 <NHSNumber>4917111072</NHSNumber>
 <StrategicReportingIdentifier>
  246216
 </StrategicReportingIdentifier>
</Identity>
<Demographics>
 <Title>Miss</Title>
 <FirstName>Alice</FirstName>
 <MiddleNames/>
 <Surname>Abbot</Surname>
 <DateOfBirth>1951-01-01</DateOfBirth>
 <Sex>F</Sex>
 <MaritalStatus/>
 <Ethnicity/>
 <MainLanguage/>
 <EnglishSpeaker>Unknown</EnglishSpeaker>
 <Address>
  <County>No Fixed Abode</County>
  <Postcode>ZZ99 3VZ</Postcode>
 </Address>
 <HomeTelephone/>
 <WorkTelephone/>
 <MobileTelephone/>
 <AlternateTelephone/>
 <EmailAddress/>
 <SMSConsent>No</SMSConsent>
 <UsualGPUserName>andrewleece2204</UsualGPUserName>
 <CareStartDate>2015-06-10</CareStartDate>
 <RegistrationType>Applied</RegistrationType>
</Demographics>
<Clinical>
 <Event DateTime="2015-06-10T15:17:30" DoneAt="Surgery" 
 DoneBy="CURCIN, Vasa (General Medical Practitioner) " 
 EventUID="f7b1080000000000" UserName="vcurcin2304">
  <Narrative>
   <Line> D: Gastro-oesophageal reflux disease (X3003)</Line>
  </Narrative>
  <Medication>
   <MedicationType>NHS medication</MedicationType>
   <Drug Name="Omeprazole 20mg dispersible gastro-resistant 
   tablets" ProductID="2007553" Scheme="Multilex"/>
   <Dose>take one daily</Dose>
   <Quantity>28 tablet</Quantity>
   <StartDate>2015-06-10</StartDate>
   <EndDate>2015-07-08</EndDate>
   <EndReason/>
  </Medication>
  <ClinicalCode>
   <Code Code="X3003" Scheme="CTV3" Term="Gastro-oesophageal 
   reflux disease"/>
  </ClinicalCode>
  <ClinicalCode>
   <Code Code="X3003" Scheme="CTV3" Term="Gastro-oesophageal 
   reflux disease"/>
   <ProblemUID>ed58000000000000</ProblemUID>
   <ProblemSeverity>Major</ProblemSeverity>
  </ClinicalCode>
 </Event>
 <Event DateTime="2015-07-14T15:31:56" DoneAt="Surgery" 
 DoneBy="CURCIN, Vasa (General Medical Practitioner) " 
 EventUID="b202680000000000" Software="CONSULT" 
 UserName="vcurcin2304">
  <Attachment>
   <FileName>testfile1.htm</FileName>
   <FileSize>0 bytes</FileSize>
   <DocumentUID>62fd200000000000</DocumentUID>
   <Comments>TESTFILE1</Comments>
 </Attachment>
 </Event>
 <Event DateTime="2015-09-24T18:09:36" DoneAt="Surgery" 
 DoneBy="CURCIN, Vasa (General Medical Practitioner) " 
 EventUID="b4f6290000000000" UserName="vcurcin2304">
  <Narrative>
   <Line>Lorem ipsum dolor sit amet, consectetur 
   adipiscing elit.</Line>
  </Narrative>
 </Event>
 <Event DateTime="2017-08-29T12:16:35" DoneAt="Surgery" 
 DoneBy="CURCIN, Vasa (General Medical Practitioner) " 
 EventUID="1bbe0d0000000000" UserName="vcurcin2304">
  <ClinicalCode>
   <Code Code="XaBVJ" Scheme="CTV3" Term="Clinical findings"/>
   <Episodicity>New</Episodicity>
   <ProblemUID>a8fa000000000000</ProblemUID>
   <ProblemSeverity>Major</ProblemSeverity>
  </ClinicalCode>
 </Event>
 <Event DateTime="2017-08-29T12:18:23" DoneAt="Surgery" 
 DoneBy="CURCIN, Vasa (General Medical Practitioner) " 
 EventUID="2bbe0d0000000000" UserName="vcurcin2304">
  <Narrative>
   <Line>H: Raised blood pressure (XM02V) D: Hypertension</Line>
  </Narrative>
  <ClinicalCode>
   <Code Code="XM02V" Scheme="CTV3" Term="Raised blood 
   pressure"/>
  </ClinicalCode>
 </Event>
 <Event DateTime="2017-08-29T12:33:22" DoneAt="Surgery" 
 DoneBy="CURCIN, Vasa (General Medical Practitioner) " 
 EventUID="4bbe0d0000000000" UserName="vcurcin2304">
  <ClinicalCode>
   <Code Code="621.." Scheme="CTV3" Term="Patient currently 
   pregnant"/>
   <ProblemUID>58fa000000000000</ProblemUID>
   <ProblemSeverity>Minor</ProblemSeverity>
   <ProblemEndDate>2017-08-29</ProblemEndDate>
  </ClinicalCode>
 </Event>
 <Event DateTime="2018-01-03T16:18:56" DoneAt="Surgery" 
 DoneBy="CHAPMAN, Martin (Administrator) "
 EventUID="3ef5ed0000000000" UserName="Martin1311">
  <ClinicalCode>
   <Code Code="XE0Ub" Scheme="CTV3" Term="Hypertension"/>
   <Episodicity>New</Episodicity>
   <ProblemUID>b8fb000000000000</ProblemUID>
   <ProblemSeverity>Major</ProblemSeverity>
  </ClinicalCode>
  <ClinicalCode>
   <Code Code="X3003" Scheme="CTV3" Term="Gastro-oesophageal 
   reflux disease"/>
   <ProblemUID>88fb000000000000</ProblemUID>
   <ProblemSeverity>Major</ProblemSeverity>
  </ClinicalCode>
 </Event>
 <Event DateTime="2018-01-03T17:02:49" DoneAt="Surgery" 
 DoneBy="CHAPMAN, Martin (Administrator) " 
 EventUID="2ef5ed0000000000" UserName="Martin1311">
  <ClinicalCode>
   <Code Code="XE0aO" Scheme="CTV3" Term="Gastro-oesophageal 
   reflux disease without oesophagitis"/>
   <ProblemUID>48fb000000000000</ProblemUID>
   <ProblemSeverity>Major</ProblemSeverity>
  </ClinicalCode>
 </Event>
 <Event DateTime="2018-01-03T17:38:25" DoneAt="Surgery" 
 DoneBy="CHAPMAN, Martin (Administrator) " 
 EventUID="09b6ed0000000000" UserName="Martin1311">
  <Medication>
   <MedicationType>NHS medication</MedicationType>
   <Drug Name="Omeprazole 20mg dispersible gastro-resistant 
   tablets" ProductID="2007553" Scheme="Multilex"/>
   <Dose>take one daily</Dose>
   <Quantity>28 tablet</Quantity>
   <StartDate>2018-01-03</StartDate>
   <EndDate>2018-01-31</EndDate>
   <EndReason/>
  </Medication>
 </Event>
</Clinical>
<NonClinical>
  <Task>
   <TaskUID>122e900000000000</TaskUID>
   <DateTime>2015-09-24T18:17:01</DateTime>
   <TaskType>Miscellaneous</TaskType>
   <Status>Not Started</Status>
   <Content>
    Thu 24 Sep 18:17 - CURCIN, Vasa Lorem ipsum dolor sit 
    amet, consectetur adipiscing elit.
   </Content>
  </Task>
  <Task>
   <TaskUID>022e900000000000</TaskUID>
   <DateTime>2015-09-24T18:16:51</DateTime>
   <TaskType>Miscellaneous</TaskType>
   <Status>Not Started</Status>
   <Content>
    Thu 24 Sep 18:16 - CURCIN, Vasa, Lorem ipsum dolor sit 
    amet, consectetur adipiscing elit.
   </Content>
  </Task>
</NonClinical>
<RecordState>Closed</RecordState>
\end{lstlisting}

\bibliographystyle{unsrt} 
\bibliography{bib}

\end{document}